# Understanding of coincidence detection in Franson-type nonlocal correlations for second-order quantum superposition

Byoung S. Ham

School of Electrical Engineering and Computer Science, Gwangju Institute of Science and Technology,
123 Chumdangwagi-ro, Buk-gu, Gwangju 61005, S. Korea
(March 15, 2022)

**Abstract**
Coincidence detection is a key technique used in nonlocal quantum-correlation measurements to test Bell's inequality violation between remotely separated local detectors. With individual uniform intensity of local measurements, the nonlocal correlation fringe is a mysterious quantum feature that cannot be achieved classically. Here, the coincidence detection is coherently investigated to understand the fundamental physics of the nonlocal correlation fringe via second-order quantum superposition between selected nonlocal measurement events. Because of the coherence feature of paired photons, the coincidence technique modifies the measurement events for the rule of thumb of indistinguishability between selected measurement bases of paired photons. This indistinguishability is quantum superposition between nonlocally detected events resulting from a selected time slot of coincidence, where coherence between individually measured photons is an absolute condition.

*Introduction*
The essential components of quantum technologies are quantum superposition and quantum entanglement [1,2]. Quantum superposition is normally understood as self-interference of a single photon in an interferometric system [3-7]. Quantum entanglement is between two correlated entities via quantum superposition with inseparable intensity product combinations [8-20]. Bell inequality violation has been intensively studied to oppose the local realism based on a hidden variable theory [8-14]. Nonlocal correlation fringe in a Franson scheme has been applied to loophole-free quantum communications, where the nonlocal correlation fringe has been left as a mysterious quantum feature because of the uniform intensity in individual and local measurements [15-20]. Although nonlocal quantum correlation has been vastly applied for various quantum technologies, including quantum key distributions [21-25], understanding the fundamental physics of the nonlocal phenomenon has been limited, where the particle nature-based interpretation does not provide a clear view of the weirdness.

In this study, a pure classical approach of coherence optics is used to investigate the mystery of nonlocal correlation fringe in a Franson scheme of Bell inequality violation. For the light source, typical entangled photon pairs from a spontaneous parametric down conversion (SPDC) process are used to analyze both local and nonlocal correlations [26,27]. For the Franson-type nonlocal correlation fringe, a general method of coincidence detection is investigated for coherence optics-caused path-basis superposition via a selected measurement process [25]. This nonlocally selected measurement process results in path-basis superposition, satisfying indistinguishability between locally measured individual event products in a specified time slot of coincidence. Thus, the coincidence detection technique plays as a major role in second-order quantum superposition between locally measured events, resulting in the nonlocal correlation fringe, otherwise satisfying a classical intensity product between local detections.

*Local randomness versus decoherence of entangled photon pairs*
In the typical Franson scheme of Fig. 1(a), an ensemble of entangled photon pairs is provided by the second-order nonlinear optics-based SPDC processes [27]. The symbol S indicates the entangled photon pair source that fires paired photons in opposite directions to satisfy the momentum conservation law. Each paired photon from S enters a corresponding noninterfering Mach-Zehnder interferometer (NMZI) comprising short (S) and long (L) paths that are simultaneously measured by both detectors in each party, Alice or Bob. For nonlocal correlation measurements, each pair of detectors in both parties is jointly measured via coincidence detection. The path-length difference $\Delta L$ $(L-S)$ of the NMZI is set to $\Delta L \gg c\Delta f^{-1}$, where $\Delta f$ is the full-width-at-half-maximum of the entangled photons, as shown in Figs. 1(b), and the c is the speed of light. Thus, the ensemble measurements of the output photons from each NMZI do not result in interference fringe. The amplitude of an arbitrary $j^{th}$ output photon from in the NMZI of Alice can be separately represented by pure coherence optics according to the wave nature of quantum mechanics:

$$E_{1j} = \frac{E_0}{2} e^{i\eta} (|S\rangle - |L\rangle e^{i\varphi_j}), \tag{1}$$

$$E_{2j} = \frac{iE_0}{2} e^{i\eta} (|S\rangle + |L\rangle e^{i\varphi_j}), \tag{2}$$

where $\eta$ is the relative phase caused by the path-length difference between S and L in the NMZI. The notations of $|S\rangle$ and $|L\rangle$ denote unit vectors of a photon state propagating along S and L paths of the NMZI, respectively. In the same way, the output photons from the NMZI of Bob can also be described as:

$$E_{3j} = \frac{E_0}{2} (|S\rangle - |L\rangle e^{i\psi_j}), \tag{3}$$

$$E_{4j} = \frac{iE_0}{2} (|S\rangle + |L\rangle e^{i\psi_j}). \tag{4}$$

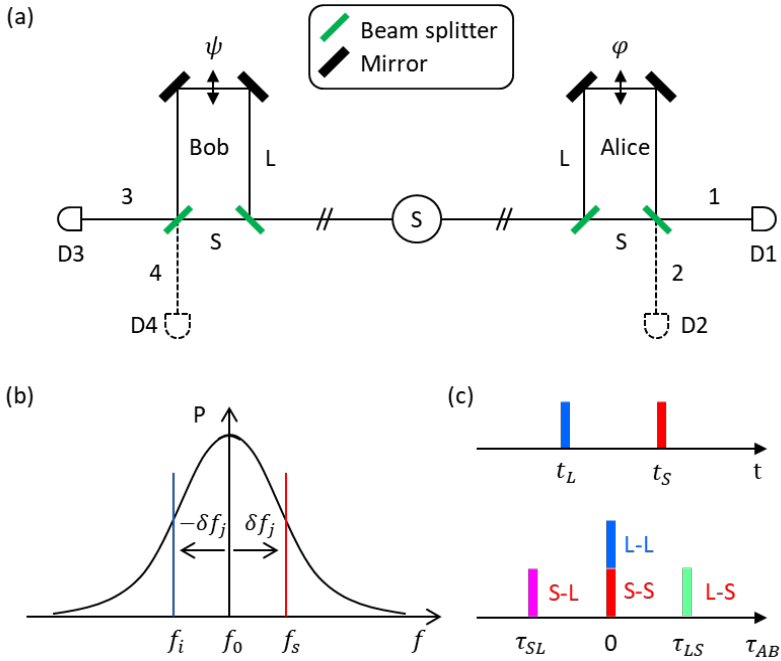

**Fig. 1. Schematic of the Franson-type nonlocal correlation.** (a) The original Franson setup. (b) Probability distribution of the frequency of the light source S in (a). (c) Schematic of Type I SPDC. $2f_0 = f_s + f_i$. S (short path) and L (long path) indicate the two paths of each unbalanced MZI. D: single photon detector. In (b), the full width at half maximum of entangled photons is $\Delta f$. The subscript k in $f_k$ indicates the signal and idler in SPDC. (c) Photon arrival time to the detectors. Upper: absolute, Lower: coincidence.

The photon intensity $I_{1j}$ of $E_{1j}$ is represented in Equation (5) assuming that each photon's coherence length $l_c$ is much longer than $\Delta L$:

$$I_{1j} = \frac{I_0}{2} (\langle S|S\rangle + \langle S|S\rangle - \langle S|L\rangle e^{i\varphi_j} + c.c.)$$

$$= \frac{I_0}{2} (1 - \cos\varphi_j). \tag{5}$$

Here, the term $\langle S|L \rangle$ is the measurement probability of coherence, resulting in an interference fringe in each local detector. This coherence term of the cosine function in Equation (5) is dependent on the coherence length $l_c$ of each entangled photon pair, where the $l_c$ is determined by the pump photon used for SPDC. As indicated by the experimental parameters [16,23], the relation of $l_c$ and $c\Delta f^{-1}$ with respect to $\Delta L$ is essential for the Franson nonlocal correlation fringe. Likewise, all other locally measured intensities are $I_{2j} = \frac{I_0}{2}(1 + cos\varphi_j)$, $I_{3j} = \frac{I_0}{2}(1 - cos\psi_j)$, and $I_{4j} = \frac{I_0}{2}(1 + cos\psi_j)$ for $l_c \gg \Delta L$.

For the wide bandwidth-distributed photon pairs in the SPDC nonlinear optical process in Fig. 1(b), the mean intensity $\langle I_1 \rangle$ can result in a uniform value without interference fringe:

$$\langle I_1 \rangle = \frac{\langle I_0 \rangle}{2N} \sum_j^N (1 + cos\varphi_j) = \frac{\langle I_0 \rangle}{2}. \qquad (6)$$

In Equation (6), $\varphi_j = \Delta f_j \tau$, where $\tau = \Delta L/c$. If $\Delta f_j \tau \gg \pi$, Equation (6) is satisfied because of the overall decoherence of the photon ensemble [16-25]. This condition can be easily satisfied with $\Delta L \gg c\Delta f^{-1}$ in each NMZI, where the usual photon bandwidth $\Delta f$ of SPDC is $>$ THz, corresponding to $<$ ps coherence time or $<$ mm coherence length. Similarly, all other mean local measurements have the same uniform intensity: $\langle I_2 \rangle = \langle I_3 \rangle = \langle I_4 \rangle = \frac{\langle I_0 \rangle}{2}$. Thus, the local measurements of uniform intensity are clearly understood using pure coherence optics, where each photon's coherence length $l_c$ must be much longer then $\Delta L$ to avoid visibility reduction. Thus, both conditions of $\Delta L \gg c\Delta f^{-1}$ and $\Delta L \ll l_c$ are satisfied for the NMZIs, resulting in the uniform local intensity. As indicated in the experimental parameters for Franson-type nonlocal correlation, these two conditions are well satisfied using a narrow bandwidth pump laser for SPDC, whose linewidth is narrow enough to be 1-10 GHz, corresponding 3-30 cm in coherence length [16,27]. The uniform local intensity has nothing to do with basis randomness in an interferometric system, but results from ensemble decoherence.

If the bandwidth condition is reversed to $\Delta f \tau \ll \pi$, then each locally measured intensities result in an interference fringe:

$$\langle I_1 \rangle = \langle I_3 \rangle = \frac{\langle I_0 \rangle}{2}(1 - cos\varphi), \qquad (7)$$

$$\langle I_2 \rangle = \langle I_4 \rangle = \frac{\langle I_0 \rangle}{2}(1 + cos\varphi). \qquad (8)$$

However, the quintessence of locally measured uniform intensities in Equations (5) and (6) is in the coincidence detection for indistinguishability of measured photons. The local fringe in Equations (7) and (8) contradicts the coincidence detection method because the S and L path cannot be separated for this photon. In SPDC, each entangled photon characteristic is governed by the pump photon, while the bandwidth of the entangled photon ensemble is governed by the SPDC process [27]. The linewidth of the pump laser is essential to determining the center frequency stability of the photon ensemble in Fig. 1(b), affecting the visibility of Bell inequality.

*Coincidence-based nonlocal correlation: second-order path-basis superposition of entangled photons*
The essence of quantum entanglement in quantum mechanics is in the nonlocal correlation between two individual measurements. Regarding bipartite particles, such as a SPDC-generated photon pair of signal and idler photons, the Bell inequality violation is achieved by a coincidence detection method between two independent and individual local measurements. However, the definition of coincidence detection rules out independence of local measurements, where mutual correlation between SPDC-generated entangled photons is coincidently tested via modified intensity products (discussed below). This is quite important to understanding how the nonlocal correlation fringe occurs. Indeed, the coincidence detection technique results in indistinguishability between measured nonlocal events for nonlocal correlation via path-basis superposition between NMZIs.

In both NMZIs of Fig. 1(a), each output photon can be represented by linear superposition of both paths of S and L, as shown in Equations (1)-(4). Thus, there are four different path-basis product combinations for nonlocal correlation measurements between NMZIs. First, the paired photons pass through only short paths, resulting in a S-S path relation. Second, the paired photons pass through only long paths, resulting in a L-L relation. Third, the paired photons pass through both short and long paths, resulting in a S-L or L-S relation. In Fig. 1(c), the top plot is for photons measured in an absolute time domain, while the bottom plot is in a coincidence time domain. In the coincidence time domain defined by $\tau_{AB}$, where $\tau_{AB} = t_A - t_B$, subscript A and B indicate Alice and Bob, respectively. For $\tau_{AB} = 0$, both S-S and L-L choices are permitted by the definition of coincidence detection, because the S-L and L-S choices are timely separated via the incoherence feature given by $\Delta L \gg \Delta f_j c$. These measurement cases of the S-L and L-S under the condition of $\Delta L \ll l_c$ must be removed from measurements of nonlocal correlation. This is a general understanding of coincidence detection for the nonlocal measurement process. Thus, this coincidence process results in a measurement modification or selection, similar to the dispersion of sunlight in a prism. The purpose of this measurement-event modification by coincidence detection is to induce indistinguishability between non-locally measured photon pairs. Here, indistinguishability is an essential condition for quantum superposition in coherence optics. For example, in Equations (7) and (8), the indistinguishability is achieved through the unknown information of a photon taking paths S and L probabilistically, resulting in the first-order intensity fringe of self-interference for a single photon [3].

Using Equations (1)-(4), the coincidently measured nonlocal correlation $R_{AB}(\tau = 0)$ between two corresponding local measurements in both parties can be described as:

$$E_{AB}^{(j)}(\tau = 0) = \frac{I_0}{4} e^{i\eta} \big(|S\rangle_A \pm |L\rangle_A e^{i\varphi_j}\big)\big(|S\rangle_B \mp |L\rangle_B e^{i\psi_j}\big),$$

$$= \frac{I_0}{4} e^{i\eta} \big(|SS\rangle_{AB} - |LL\rangle_{AB} e^{i(\varphi_j + \psi_j)}\big), \tag{9}$$

where both terms of $|SL\rangle_{AB}$ and $|LS\rangle_{AB}$ are ruled out by the definition of coincidence detection. Here, $R_{AB}(\tau = 0) = E_{AB} E_{AB}^*$ and $E_{AB} = E_A E_B$. The $E_{AB}$ can never be achieved without the coincidence detection, resulting in the quintessence of nonlocal correlation. Thus, the following nonlocal fringe for each photon pair is achieved:

$$R_{AB}^{(j)}(\tau = 0) = \frac{I_0^2}{8} \big(1 - \cos(\varphi_j + \psi_j)\big), \tag{10}$$

where $\varphi_j = \varphi + \delta f_j$ and $\psi_j = \varphi - \delta f_j$ due to the symmetric detuning of SPDC-generated entangled photon pairs as shown in Fig. 1(b), satisfying energy conservation law in SPDC [27]. Unlike local measurements, Equation (10) is independent of $\delta f_j$ of the $j^{th}$ photon pair. Thus, the time averaged (mean) nonlocal correlation is only dependent on both NMZI phase controls:

$$\langle R_{AB}(\tau = 0) \rangle = \frac{\langle I_0^2 \rangle}{8} \big(1 - \cos(\varphi + \psi)\big). \tag{11}$$

Assuming well stabilized NMZIs from environmental noises such as air turbulence, temperature fluctuations, and mechanical vibrations, thus the mysterious nonlocal fringe in the Franson scheme is successfully analyzed using pure coherence optics via coincidence detection method. Here, the coincidence-resulting path-basis superposition in Equation (9) plays a key role for the joint phase control of $\cos(\varphi + \psi)$ in Equation (11) via the second-order quantum superposition of entangled photon pairs, resulting in the doubled fringe pattern compared with the first-order quantum superposition in Equations (5) for a synchronized phase control with $\varphi = \psi$.

For the nonlocal fringe in Equation (11), the quintessence is coincidence detection-caused superposition between path choices of S-S and L-L, resulting in indistinguishability of the path-basis choice for the paired photons. This relation is exactly the same as quantum superposition in a single MZI between two paths of S and L in Equations (1)-(4) for a single photon. As shown in the photonic de Broglie waves for N=2 [28], the doubled fringe results from Equation (11) compared with Equations (1)-(4), if both phases were synchronized. This is the essence of higher-order quantum superposition, where the superposition order can be for photon numbers for a fixed MZI [28-31] or MZI numbers [32] for a fixed photon. Equation (11) is related to the number of photons rather than the number of MZIs. Thus, macroscopic nonlocal correlation can be possible in coherence optics. For the nonlocal fringe in Equation (11), the coherence length $l_C$ of each SPDC-generated photon pair must be longer than the path-length difference $\Delta L$ of NMZI. The coincidence-caused nonlocal fringe via second-order quantum superposition is the main result of this paper for understanding nonlocal correlation.

*Discussion*

If the non-coincidence terms (S-L and L-S path choices) are not eliminated for the nonlocal correlation measurements, what happens to the nonlocal fringe in Equation (10)? This question is essential to understanding the nonlocal quantum feature in terms of quantum superposition between basis choices of measurement entities. Simply, Equation (9) is not effective, resulting in no $\cos(\varphi_j + \psi_j)$ term as in Equation (10) due to the locally separated intensities, resulting in no $E_{AB}(=E_A E_B)$ term. Thus, the intensity correlation without a coincidence detection technique just follows a classical manner of separable products, $R_{AB}^{(j)} = \frac{I_0^2}{8}(1 \pm \cos\varphi_j)(1 \mp \cos\psi_j)$, resulting in $\langle R_{AB} \rangle = \sum_j^N R_{AB}^{(j)} = \frac{I_0^2}{4}$ due to the wide bandwidth $\Delta f$. In other words, the nonlocal correlation fringe in a Franson scheme is due to measurement-event modifications (or selections) via coincidence detection, resulting in path-basis superposition of paired photons. This is the most important analytic result in this coherence interpretation of the Franson-type nonlocal correlation. As noted by Feynman, the only mysterious feature in quantum mechanics is quantum superposition [33]. The resulting second-order quantum superposition in Equation (9) may be implemented macroscopically using coherent light (presented elsewhere).

According to the photon characteristics of SPDC-generated entangled photon pairs [27], the locally measured uniform intensity in each party is a direct result of decoherence by wide bandwidth-distributed ensemble photons in NMZI. Although the locally measured uniform intensity has nothing to do with the path-basis randomness in Equation (9); it is a necessary condition for the coincidence detection to separate S-S and S-L via bandwidth-caused decoherence. The nonlocal fringe is rooted in the measurement basis randomness between S-S and L-L path choices, resulting in the path-basis superposition in a second-order form. Without coincidence detection-caused measurement-basis randomness (indistinguishability) between S-S and L-L in nonlocal measurements, the phase-added cosine term in Equation (11) is not possible. Unlike the common understanding, the measured Franson-type nonlocal correlation fringe requires coherence between measured entities of paired photons for measurement-basis superposition. This coincidence-induced second-order quantum superposition does not necessarily belong to the particle nature of photons. Instead, it is a direct result of the wave nature of photons, i.e., coherence optics.

*Conclusion*

The fundamental physics of nonlocal correlation in a Franson scheme was analyzed using pure coherence optics, where the nonlocal fringe resulted from filtered measurement events via a coincidence detection process. Locally measured uniform intensity in each detector was due to decoherence among wide bandwidth-distributed entangled photon pairs via many-wave interference, whereas each photon pair must be perfectly coherent in the seemingly noninterfering MZIs of the Franson scheme. This locally measured uniform intensity in each detector is an essential condition for coincidence detection in nonlocal measurements to avoid unnecessary measurements. Coincidence detection-caused filtered events were to satisfy the path-basis second-order quantum superposition of individually measured paired photons, resulting in indistinguishability between two events in a coincidence time slot. For this event-choice randomness, coherence between measured photon pairs was a necessary condition. As a result, the nonlocal correlation fringe was successfully analyzed as a direct result of coherence optics based on symmetrically detuned SPDC-generated photon pairs via coincidence detection. Thus, the mysterious quantum feature of nonlocal correlation fringe in a Franson scheme was successfully understood as a quantum feature of measurement indistinguishability via coincidence detection. This coincidence detection created quantum superposition between locally measured joint evens in an inseparable product manner. Without coincidence detection-caused measurement modification, such nonlocal quantum superposition is impossible, resulting in separable products between locally measured events.

**Acknowledgment**
This work was supported by GIST via GRI 2022 and the ICT R&D program of MSIT/IITP (2021-0-01810), Development of Elemental Technologies for Ultra-secure Quantum Internet.